\documentclass[aps,pra,twocolumn,superscriptaddress,showpacs,10pt]{revtex4-1}
\usepackage{amsmath,amssymb,graphicx,color}
\usepackage{epstopdf}

\begin{document}
\title{Suppression of quantum dissipation: A cooperative effect of quantum squeezing and quantum measurement}

\author{Yi-Ming Xia}
\affiliation{College of Physics and Electronic Science and Hubei Key Laboratory of Photoelectric Materials and Devices, Hubei Normal University, Huangshi 435002, China}

\author{Yi-Fei Wang}
\affiliation{College of Physics and Electronic Science and Hubei Key Laboratory of Photoelectric Materials and Devices, Hubei Normal University, Huangshi 435002, China}

\author{Xiao-Yun Zhang}
\affiliation{College of Physics and Electronic Science and Hubei Key Laboratory of Photoelectric Materials and Devices, Hubei Normal University, Huangshi 435002, China}

\author{Hai-Chao Li}
\altaffiliation{hcl2007@foxmail.com}
\affiliation{College of Physics and Electronic Science and Hubei Key Laboratory of Photoelectric Materials and Devices, Hubei Normal University, Huangshi 435002, China}

\author{Wei Xiong}
\altaffiliation{xiongweiphys@wzu.edu.cn}
\affiliation{Department of Physics, Wenzhou University, Wenzhou 325035, China}

\begin{abstract}
The ability to isolate a quantum system from its environment is of fundamental interest and importance in optical quantum science and technology. Here we propose an experimentally feasible scheme for beating environment-induced dissipation in an open two-level system coupled to a parametrically driven cavity. The mechanism relies on a novel cooperation between light-matter coupling enhancement and frequent measurements. We demonstrate that, in the presence of the cooperation, the system dynamics can be completely dominated by the effective system-cavity interaction and the dissipative effects from the system-environment coupling can be surprisingly ignored. This work provides a generic method of dissipation suppression in a variety of quantum mechanical platforms, including natural atoms and superconducting circuits.
\end{abstract}

\maketitle

Coherent interaction between a two-level atom and a quantized single-mode electromagnetic field describes the fundamental mechanism of light-matter interaction at the quantum level. Typically, this interaction can be captured by the well-known Jaynes-Cummings model~\cite{Jaynes,Shore} which plays a crucial role in boosting the development of quantum optics\cite{Scully} and quantum information science~\cite{Nielsen}. In particular, the Jaynes-Cummings model predicted the existence of periodic collapse and revival of Rabi oscillations~\cite{Eberly}, which can serve as a paradigm of exploring fascinating physics in atomic cavity quantum electrodynamics (QED)~\cite{Walther,Reiserer}. Over the years, the Jaynes-Cummings model and its generalizations have been used to demonstrate numerous interesting quantum phenomena and practical quantum technologies, such as electromagnetically induced transparency~\cite{Boller,Fleischhauer}, quantum phase transition~\cite{Hwang,Li,Zhang}, and quantum entanglement~\cite{Karlsson,Horodecki}. Apart from naturally occurring atoms, this model nowadays can be implemented in many artificially engineered quantum systems, for example superconducting quantum circuits~\cite{Blais,Wallraff,Gu}.

Beyond a closed-system description in the Jaynes-Cummings model, every realistic quantum system is unavoidably coupled to its surrounding environment and in nature, should be deemed to be an open system~\cite{Breuer}. In this scenario, the environment tends to introduce irreversible dissipation processes triggering a nonunitary time evolution for the system. As a consequence, the dissipation makes the dynamics within the system decoherent in a long-time scale, which is a long-standing serious obstacle to a large number of coherence-based quantum effects and applications, ranging from quantum computation~\cite{Xu} to quantum metrology~\cite{Smirne} and quantum biology~\cite{Chin}. Hence, a natural open question that we address here arises: How can one develop an accessible method to suppress or even neutralize the environment-induced detrimental effects on the quantum system? Interestingly, considerable effort has been devoted to the study of dealing with this question, such as quantum error correction~\cite{Shor}, dynamical decoupling~\cite{Viola}, and distance-based coherence freezing~\cite{Bromley}.

Here we present an alternative method to suppress dissipation in an open two-level system interacting with a quantized cavity field. Given the coexistence and competition between coherent system-cavity coupling and incoherent system-environment coupling, one of our strategies is to enhance the coherent coupling and makes it work as a dominant role in the competition. Especially, for a weak-coupling cavity-QED system, it is highly desirable to push atom-field interaction into a strong-coupling regime and this coupling enhancement can be realized by using quantum squeezing~\cite{Leroux,Qin}. Meanwhile, the other strategy is to remove the long-time accumulation from the system-environment coupling via frequently repeated measurements~\cite{Misra,Erez}.
\begin{figure}[htbp]
\centerline{
\includegraphics[width=0.9\columnwidth]{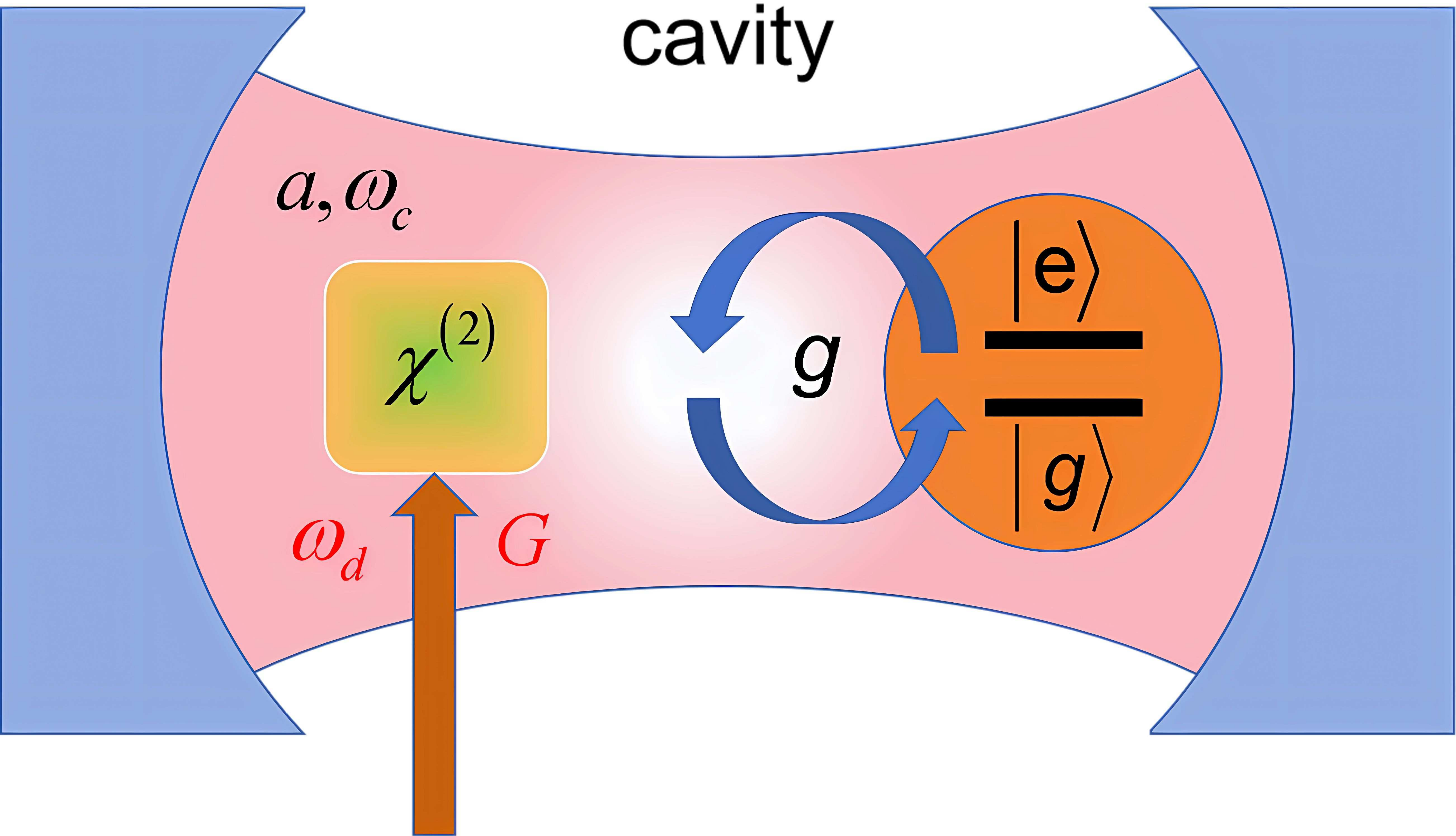}}
\caption{(Color online) Schematic of a parametrically driven cavity interacting with a two-level system. The parametric drive on the cavity is generated by pumping a $\chi^{(2)}$ nonlinear medium with amplitude $G$ and frequency $\omega_{d}$. Here $|e\rangle$ ($|g\rangle$) is the ground (excited) state of the two-level system, $a$ is the annihilation operator of the cavity mode with frequency $\omega_{c}$, and $g$ is the strength of the system-cavity interaction.
}
\label{fig.1}
\end{figure}
We show that, with the help of a cooperative mechanism between quantum squeezing and frequent measurements, the dynamical time evolution for the open system is totally governed by the coherent coupling and hence, the environmental influence can be safely neglected during the whole evolution.
Our proposal can be implemented in a wide range of physical systems and, in particular, two examples in natural atoms and superconducting qubits are displayed.

We consider a parametrically driven Jaynes-Cummings model where a single-mode cavity interacts with a two-level system coupled to its dissipative environment. The cavity contains a $\chi^{(2)}$ nonlinear medium strongly pumped by an external field, which generates a parametric drive on the cavity mode, as shown in Fig~\ref{fig.1}. This model can be described by the Hamiltonian ($\hbar=1$)
\begin{align}\label{1}
H=&\frac{1}{2}\omega_{q}\sigma_{z}+\omega_{c}a^{\dag}a+g(\sigma_{+}a+\sigma_{-}a^{\dag})\nonumber \\
&+\sum_{k}\omega_{k}b_{k}^{\dag}b_{k}+\sum_{k}f_{k}(\sigma{+}b_{k}+\sigma_{-}b_{k}^{\dag})\nonumber \\
&-\frac{G}{2}({a^{\dag}}^2e^{-i\omega_{d}t}+a^2e^{i\omega_{d}t}),
\end{align}
where $\sigma_{z}$ is the Pauli matrix of the system with frequency $\omega_{q}$, $\sigma_{\pm}$ are the raising and lowering operators between excited state $|e\rangle$ and ground state $|g\rangle$, $a$ and $a^{\dag}$ ($b_{k}$ and $b_{k}^{\dag}$) are the annihilation and creation operators of the cavity mode ($k$th mode of the environment) with frequency $\omega_{c}$ ($\omega_{k}$), and $g$ ($f_{k}$) is the system-cavity (system-environment) coupling strength. The last term in the Hamiltonian denotes the parametric drive with amplitude $G$ and frequency $\omega_{d}$. Without the dissipation and the drive, Eq.~(\ref{1}) would reduce to the standard Jaynes-Cummings model.

Working in a frame rotating with $\omega_{d}/2$, the Hamiltonian can be written as
\begin{align}\label{2}
H=&\frac{1}{2}\Delta_{q}\sigma_{z}+\Delta_{c}a^\dag a+g(\sigma_{+}a+\sigma_{-}a^\dag)\nonumber \\
&+\sum_{k}\Delta_{k}b_{k}^\dag b_{k}+\sum_{k}f_{k}(\sigma_{+}b_{k}+\sigma_{-}b_{k}^\dag)\nonumber \\
&-\frac{G}{2}(a^{\dag2}+a^2),
\end{align}
where $\Delta_{q,c,k}=\omega_{q,c,k}-\omega_{d}/2$ are the detunings of the two-level system, the cavity field, and the $k$th mode of the environment, respectively. In order to realize coherent coupling amplification between the system and the cavity mode, we introduce a preferred squeezed cavity mode $a_{s}$ by utilizing a squeezing transformation $a=\textrm{cosh}(r_s)a_s+\textrm{sinh}(r_s)a_s^\dagger$ with a squeezing parameter $r_{s}=(1/4)\ln[(\Delta_{c}+G)/(\Delta_{c}-G)]$. After making this substitution into Eq.~(\ref{2}), the Hamiltonian under a rotating-wave approximation becomes
\begin{align}\label{3}
H_s=&\frac{1}{2}\Delta_{q}\sigma_{z}+\Delta_{s}a_s^\dag a_s+g_s(\sigma_{+}a_s+\sigma_{-}a_s^\dag)\nonumber \\
&+\sum_{k}\Delta_{k}b_{k}^\dag b_{k}+\sum_{k}f_{k}(\sigma_{+}b_{k}+\sigma_{-}b_{k}^\dag),
\end{align}
where $\Delta_{s}=\sqrt{\Delta_{c}^2-G^2}$ is the frequency of the squeezed cavity mode and $g_s=g\textrm{cosh}(r_{s})$ is the interaction strength of the two-level system with the squeezed mode. When the squeezing parameter is tuned to satisfy $r_{s}\gg1$ easily achieved at the threshold $\Delta_{c}\rightarrow{G}$, the system-cavity coupling in the squeezed frame reduces to $g_s={g}\exp(r_{s})/2$. As a result, we obtain an exponentially enhanced atom-cavity coupling via the additional parametric drive, which can be several orders of magnitude larger than the original coupling. For example, when $r_{s}=10$, we have $g_s\simeq 2.2\times10^{4}{g}$.

We now study the quantum dynamics with frequent measurements in the squeezed frame. We assume that at the initial time t=0 the system is prepared in the excited state $|e\rangle$, and the squeezed cavity mode and all modes of the environment are in the respective vacuum states $|0\rangle$ and $|\{0\}\rangle$. In this case, the wave function takes the form
\begin{align}\label{4}
|\Psi(t)\rangle=&\alpha(t)|e,0,\{0\}\rangle+\beta(t)|g,1,\{0\}\rangle\nonumber \\
&+\sum_{k}\gamma_{k}(t)|g,0,1_{k}\rangle,
\end{align}
where $\alpha(0)=1$ and $|1_{k}\rangle\equiv|0_{1},0_{2},...,1_{k},...\rangle$ indicates single quantum excitation for the $k$th mode of the environment. It is well known that the dynamics of a quantum system can be described by the equations of motion for the probability amplitudes. According to the Schr\"{o}dinger equation $i\partial_{t}|\Psi(t)\rangle=H_{s}|\Psi(t)\rangle$, we obtain
\begin{align}\label{5}
\frac{d\widetilde{\alpha}(t)}{dt}&=-ig_s\widetilde{\beta}(t)\text{exp}[-i(\Delta_{s}-\Delta_{q})t]-i\sum_{k}f_{k}\widetilde{\gamma}_{k}(t)\nonumber\\
&\quad\times\text{exp}[-i(\Delta_{k}-\Delta_{q})t],\nonumber\\
\frac{d\widetilde{\beta}(t)}{dt}&=-ig_s\widetilde{\alpha}(t)\text{exp}[i(\Delta_{s}-\Delta_{q})t],\nonumber\\
\frac{d\widetilde{\gamma}_{k}(t)}{dt}&=-if_{k}\widetilde{\alpha}(t)\text{exp}[i(\Delta_{k}-\Delta_{q})t],
\end{align}
by making the transformations
\begin{align}\label{6}
\alpha(t)&=\widetilde{\alpha}(t)\text{exp}\left(-\frac{i}{2}\Delta_{q}t\right),\nonumber\\
\beta(t)&=\widetilde{\beta}(t)\text{exp}\left[-i\left(\Delta_{s}-\frac{1}{2}\Delta_{q}\right)t\right],\nonumber\\
\gamma_{k}(t)&=\widetilde{\gamma}_{k}(t)\text{exp}\left[-i\left(\Delta_{k}-\frac{1}{2}\Delta_{q}\right)t\right].
\end{align}
Given that frequently projective measurements are used to interrupt the time evolution (i.e., counteracting the long-time accumulation of environment-induced dissipation), it is enough to obtain the short-time dynamics and so Eq.~(\ref{5}) can be iteratively solved~\cite{Zhangj,Kofman}. After $n$ successive measurements with a sufficiently small interval $\tau$, the probability of remaining the initial state reads
\begin{align}\label{7}
P(t=n\tau)&=|\alpha(\tau)|^{2n}=\text{exp}[-\Gamma(\tau)t],
\end{align}
where $\Gamma(\tau)$ is the measurement-induced decay rate. Due to the presence of both coherent system-cavity coupling and incoherent system-environment coupling, the effective decay rate $\Gamma(\tau)$ can be divided into two independent subexpressions
\begin{align}\label{8}
\Gamma(\tau)&=\Gamma_{c}(\tau)+\Gamma_{e}(\tau)\nonumber \\
&=g_s^{2}\tau+2\pi\int^{\infty}_{0}{d\omega}G(\omega)F(\omega),
\end{align}
where $\Gamma_{e}(\tau)$ is an overlap integral of a bath spectral density
\begin{align}\label{9}
G(\omega)&=\sum_{k}f_{k}^{2}\delta(\omega-\omega_{k})
\end{align}
and a filter function generated by frequent measurements
\begin{align}\label{10}
F(\omega)&=\frac{\tau}{2\pi}\text{sinc}^{2}\left[\frac{(\omega-\omega_{q})\tau}{2}\right].
\end{align}
As a comparison, the decay rate without the parametric drive is also given by
\begin{align}\label{11}
\Gamma^{\textrm{wo}}(\tau)&=\Gamma^{\textrm{wo}}_{c}(\tau)+\Gamma^{\textrm{wo}}_{e}(\tau)\nonumber \\
&=g^{2}\tau+2\pi\int^{\infty}_{0}{d\omega}G(\omega)F(\omega).
\end{align}
Note that, for the system-environment coupling-induced part, $\Gamma_{e}(\tau)$ in Eq.~(\ref{8}) is entirely identical with $\Gamma^{\textrm{wo}}_{e}(\tau)$ in Eq.~(\ref{11}) due to the relation $\Delta_{k}-\Delta_{q}=\omega_{k}-\omega_{q}$. For convenience later, we use uniformly the notation $\Gamma_{e}(\tau)$. Obviously, the coexistence between the cavity mode and the environment can result in competitive measurement dynamics. In a sense, the coherent interaction can dominate more easily the competitive dynamics when it becomes more strong. Consequently, whether or not the parametric drive exists can have a highly significant impact on the competition process, as will be demonstrated below.

The above calculation is universally suitable for the interaction of a quantized single-mode field with a dissipative two-level system and such an interaction can be realized in a series of quantum mechanical architectures, including naturally occurring and artificially designed quantum systems. Here our discussion is devoted to exploring natural atoms and superconducting qubits.

\begin{figure}[htbp]
\centerline{
\includegraphics[width=0.99\columnwidth]{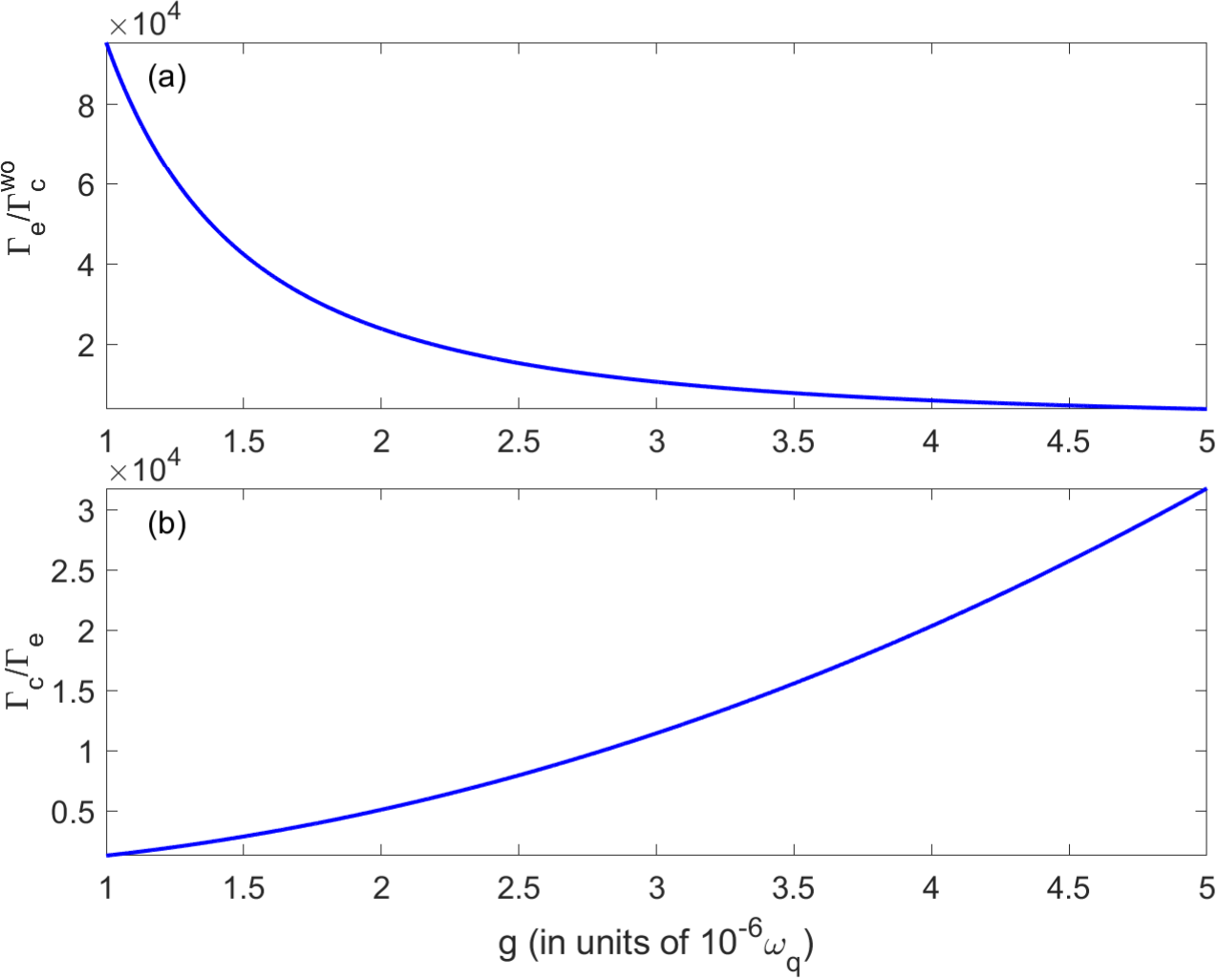}}
\caption{(Color online) (a) Renormalized value $\Gamma_{e}/\Gamma^{\textrm{wo}}_{c}$ of the hydrogen atom as a function of the coupling strength $g$. (b) Renormalized value $\Gamma_{c}/\Gamma_{e}$ under the developed cooperative mechanism with the squeezing parameter $r_{s}=10$.
}
\label{fig.2}
\end{figure}

We consider the 2\emph{P}--1\emph{S} transition of the hydrogen atom~\cite{Facchi1}with frequency $\omega_{q}=1.55\times10^{16}$rad/s and spectral density function
\begin{align}\label{12}
G(\omega)&=\frac{\eta\omega}{[1+(\omega/\omega_{s})^2]^4},
\end{align}
where $\eta=6.4\times10^{-9}$ is the dimensionless constant and $\omega_{s}=550 \omega_{q}$ is the cutoff frequency. We emphasize that in atomic cavity QED, the experimentally achievable ratio between the coupling strength $g$ and the cavity-mode frequency $\omega_c$ is no greater than $10^{-6}$~\cite{Forn}. According to Eq.~(\ref{11}), we plot the relative value $\Gamma_{e}/\Gamma^{\textrm{wo}}_{c}$ for the hydrogen atom with a typical measurement interval $\omega_{q}\tau=1$, as seen in Fig.~\ref{fig.2}(a). From the picture, we see that the quantity obtained from the cavity mode can dramatically be four orders of magnitude smaller than that of the environment. This clearly illustrates that the environment controls the system dynamics even if the system-cavity interaction exists.

In order to suppress environment-induced effects on the hydrogen atom, a feasible pathway is to enhance the atom-cavity coupling strength, which has been achieved by the quantum squeezing method. As an example, when the squeezing parameter is set to $r_{s}=10$, the ratio $\Gamma_{c}/\Gamma_{e}$ in Eq.~(\ref{8}) is presented as a function of the original strength $g$, as shown in Fig.~\ref{fig.2}(b). Obviously, the dynamical evolution in the squeezed frame can be perfectly determined by the system-cavity coupling and, in essence, the environmental role does not at all work because of $\Gamma_{c}/\Gamma_{e}\sim10^4$. Thus, by creating a cooperative mechanism between coupling enhancement and frequent measurements, we demonstrate an efficient method of dissipation suppression in an open quantum system.

In parallel to cavity QED based on natural atoms, coherent coupling between superconducting artificial atoms and microwave photons in circuit QED~\cite{You,Gux,Blais1}can be also described by the Jaynes-Cummings model.
Typically, superconducting quantum circuits can work in the strong coupling regime with $g/\omega_{q}\sim10^{-3}$ and superconducting artificial atoms are subjected to intrinsic low-frequency noise with spectral density~\cite{Cao}
\begin{align}\label{13}
G(\omega)&=\frac{2\chi\omega}{(\omega/\omega_{q})^2+(\lambda/\omega_{q})^2},
\end{align}
where $\chi$ is a dimensionless coupling strength and $\lambda$ denotes an energy lower than the energy spacing $\omega_{q}$. In terms of the cooperative mechanism developed above, we have $\Gamma_{c}/\Gamma_{e}\simeq2.5\times10^{5}$ and so the dissipative effects induced by the environment can be significantly suppressed in superconducting circuits. Without the cooperation, the environment would dominate the dynamics again due to $\Gamma_{e}/\Gamma^{\textrm{wo}}_{c}\simeq4.9\times10^{2}$. The parameters used are $\chi=10^{-4}$, $\lambda=0.05\omega_{q}$, $g=10^{-3}\omega_{q}$, $r_{s}=10$, and $\omega_{q}\tau=1$.

In summary, we have developed an accessible method of suppressing environment-induced dissipation in a generalized Jaynes-Cummings model describing the interaction of a parametrically driven cavity with an open two-level system. We have shown that, based on a cooperative mechanism of quantum squeezing and frequent measurements, the quantum dynamics can be perfectly governed by coherent light-matter interaction. Our generic scheme against dissipation can be carried out in various quantum optical architectures, such as natural atoms, superconducting circuits, and hybrid quantum systems~\cite{Xiang,Lachance}. Moreover, this work may open up a promising route to exploiting dissipation-free quantum technologies even in the presence of a dissipative environment.

This work was partially supported by the Natural Science Foundation of Hubei Province of China under Grant No. 2022CFB509, the National Natural Science Foundation of China under Grant No. 11904201, and the Natural Science Foundation of Zhejiang Province of China under Grant No. LY24A040004.

\end{document}